\documentclass[iop]{emulateapj}

\usepackage{natbib,graphicx,amsmath}
\usepackage{apjfonts}
\def\mathbi#1{\textbf{\em #1}}

\shorttitle{RELATIVISTIC NON-LINEAR POWER SPECTRA}
\shortauthors{JEONG ET AL}

\begin{document}

\title{General relativistic effects on non-linear power spectra}

\author{Donghui Jeong}
\affil{California Institute of Technology, Pasadena, CA 91125-1700, USA}
\email{djeong@tapir.caltech.edu}
\author{Jinn-Ouk Gong\altaffilmark{1}}
\affil{Instituut-Lorentz for Theoretical Physics, Universiteit Leiden, 2333 CA Leiden, The Netherlands}
\email{jgong@lorentz.leidenuniv.nl}
\author{Hyerim Noh}
\affil{Korea Astronomy and Space Science Institute, Daejeon 305-348, Republic of Korea}
\email{hr@kasi.re.kr}
\and
\author{Jai-chan Hwang}
\affil{Department of Astronomy and Atmospheric Sciences, Kyungpook National University, Daegu 702-701, Republic of Korea}
\email{jchan@knu.ac.kr}
\altaffiltext{1}{Present address: Theory Division, CERN, CH-1211 Gen\`eve 23, Switzerland}

\begin{abstract}

Non-linear nature of Einstein equation introduces genuine
relativistic higher order corrections to the usual Newtonian fluid
equations describing the evolution of cosmological perturbations.
We study the effect of such novel non-linearities on
the next-to-leading order matter and velocity power spectra for the case of
pressureless, irrotational fluid in a flat Friedmann background. We
find that pure general relativistic corrections are negligibly small
over all scales. Our result guarantees that, in the current paradigm of
standard cosmology, one can safely use Newtonian cosmology even in non-linear regimes.

\end{abstract}


\section{Introduction}

Large scale structure (LSS) of the universe is a powerful probe to
study the nature of cosmological density perturbations and to
extract cosmological parameters~\citep{peebles:1980}. Combined with the
anisotropy of the cosmic microwave background (CMB), most of the
cosmological parameters are currently constrained within a few
percent accuracy or even better~\citep{Komatsu/etal:prep}. To continue
our success in cosmology with CMB and LSS, it is crucial to predict
the power spectra from theory accurately.
While the temperature fluctuation in the
CMB is as small as $\delta{T}/T \sim 10^{-5}$~\citep{smoot/etal:1992} so that linear
perturbation theory is able to provide necessary accuracy, we have
larger degree of non-linearities in LSS. We must take into account
non-linearities of LSS properly to predict the power spectrum
accurate enough for precision cosmology at a level similar to the
CMB~\citep{jeong/komatsu:2006,jeong/komatsu:2009a}.

Most studies on LSS, however, have been based on Newtonian gravity,
especially those including non-linear
perturbations~\citep{vishniac:1983,goroff/etal:1986,makino/sasaki/suto:1992,fry:1994,bernardeau/etal:2002}. This approach has to be
justified a posteriori by comparing the result against fully general
relativistic one. For example, in \citet{noh/hwang:2004}, it is shown
that the Newtonian hydrodynamic equations up to second order
{\it coincide exactly} with the
relativistic ones~in the zero pressure case, after
appropriately identifying hydrodynamical variables with
gauge-invariant combinations of relativistic perturbation variables.
Thus, compared to the Newtonian hydrodynamic equations which are closed at
second order, any higher order contributions are originated from purely general
relativistic effects~\citep{hwang/noh:2005b}. A consistent expansion of
density fluctuation tells us that the leading non-linear
contributions to the power spectrum include third order
perturbations~\citep{noh/hwang:2008,noh/jeong/hwang:2009}. Thus non-linear density
power spectrum naturally include pure general relativistic effects,
which may have important implications as N-body simulations are
becoming larger and larger to reach the horizon
scale~\citep{kim/park/gott/dubinski:2009}.

In this note, we examine the general relativistic effects on
the power spectra of matter density fluctuations and peculiar velocity by including
leading non-vanishing non-linear contributions. Our aim is to answer
the question whether pure general relativistic effects can give rise
to any cosmologically observable consequences. To our surprise, we
find that the Newtonian terms in these power spectra are
absolutely dominating over {\em all} relevant cosmological scales,
even outside the horizon. Although the result sounds simple and
pleasant, this is still a non-trivial result, because in the context
of cosmology Newtonian gravity is {\em incomplete}: there is no
concept of horizon, the propagation speed of an action at one point
is infinite, and so on.

This note is outlined as follows. In Section~\ref{sec:formalism},
we present the formalism to set up
the equations to solve, and give
the solutions up to third order. In Section~\ref{sec:spectrum},
we compute the matter and velocity power spectra including next-to-leading
non-linear corrections which include genuine general relativistic
effects. In Section~\ref{sec:conclusions} we conclude.

\section{Equations and solutions}
\label{sec:formalism}

We consider Einstein-de Sitter universe, i.e. a flat universe
dominated by pressureless, irrotational matter, and consider only
the scalar perturbations. We work in the temporal comoving
gauge where ${T}^0{}_i = 0$ to all perturbation orders, with
$i$ being a spatial index. As this temporal gauge condition,
together with our unique spatial gauge condition~\citep{bardeen:1988}
$g_{ij}=a^2(1+2\varphi)\delta_{ij}$, fixes the gauge degrees of
freedom completely, all the resulting perturbation variables can be
equivalently regarded as fully gauge invariant, both spatially and
temporally. This statement is valid in all perturbation
orders~\citep{noh/hwang:2004}.

The Arnowitt-Deser-Misner formulation~\citep{2008GReGr..40.1997A} is
convenient in our case~\citep{1980PhRvD..22.1882B}. The comoving
gauge condition imply the momentum density vanishes, i.e. ${J}_i
\equiv N T^0{}_i = 0$, with $N$ being the lapse function. The
pressureless condition implies ${S}_{ij} \equiv T_{ij}= 0$.
Therefore the momentum conservation equation gives $N_{,i} = 0$,
thus the lapse function $N$ is uniform. The energy and momentum
conservation equations and the trace part of the propagation
equation then become~\citep{1980PhRvD..22.1882B}
\begin{align}
& {E}_{,0} - N^i{E}_{,i} = NK {E} \, ,
\\
& \overline{K}^j{}_{i|j} - \frac{2}{3}K_{|i} =  0 \, ,
\\
& K_{,0} - N^iK_{,i} = N \left( \frac{1}{3}K^2 +
\overline{K}^i{}_j\overline{K}^j{}_i + 4\pi G{E} - \Lambda \right)
\, ,
\end{align}
where $N_i$ is the shift vector, $K \equiv K^i{}_i$ is the trace of
the extrinsic curvature tensor $K_{ij}$, $\Lambda$ is the
cosmological constant, ${E} \equiv N^2 T^{00}$ is the energy
density, an overbar denotes the traceless part, and a vertical bar
denotes a covariant derivative with respect to $\gamma_{ij}$. These
are the complete equations we need in our non-linear perturbations,
and are valid in fully non-linear situation.

We introduce the density and the velocity fluctuations as $E \equiv \rho(t) + \delta\rho(t,{\mathbi x})$ and $K \equiv 3 H - \theta(t,\mathbi{x})$ with $\theta(t,\mathbi{x}) \equiv a^{-1}\nabla\cdot\mathbi{u}(t,\mathbi{x})$, with $a$ being the cosmic scale factor.
We can identify $\delta \rho(t,{\mathbi x})$ and $\mathbi{u}(t,\mathbi{x})$ as the Newtonian density and velocity perturbation variables respectively, because the relativistic equations coincide exactly with the corresponding Newtonian hydrodynamic equations up to second order.

From the above equations we can derive the hydrodynamic equations of density fluctuation $\delta(t,{\mathbi x}) \equiv \delta\rho(t,{\mathbi x})/\rho(t)$
and velocity gradient $\theta(t,\mathbi{x})$
to the third order~\citep{hwang/noh:2005b}. The relativistic
continuity and Euler equations are found to be
\begin{align}
\label{delta-dot-eq}
& \frac{\partial \delta}{\partial t} + \frac{1}{a}\nabla\cdot\mathbi{u} =
-\frac{1}{a}\nabla\cdot(\delta\mathbi{u}) + \frac{1}{a} \left[
2\varphi \mathbi{u} - \nabla \left( \Delta^{-1}X_2 \right)
\right]\cdot(\nabla\delta) \, ,
\\
\label{u-dot-eq}
& \frac{1}{a} \nabla \cdot \left( \frac{\partial
\mathbi{u}}{\partial t} + H \mathbi{u} \right) + 4 \pi G \rho
\delta = -\frac{1}{a^2} \nabla\cdot \left[ (\mathbi{u}\cdot\nabla)\mathbi{u}
\right]
\nonumber\\
& \;\;  + \frac{4}{a^2}\nabla\cdot
\left\{ \varphi \left[ (\mathbi{u}\cdot\nabla)\mathbi{u} -
\frac{1}{3}(\nabla\cdot\mathbi{u})\mathbi{u} \right] \right\}
- \frac{2}{3a^2}\varphi
(\mathbi{u}\cdot\nabla)(\nabla\cdot\mathbi{u})
\nonumber \\
& \;\;
-\frac{1}{a^2} \Delta\left[ (\mathbi{u}\cdot\nabla)\Delta^{-1}X_2
\right]
 +\frac{1}{a^2} (\mathbi{u}\cdot\nabla)X_2 +
\frac{2}{3a^2}X_2(\nabla\cdot\mathbi{u}) \, ,
\end{align}
where $\varphi$ and $X_2$ are the linear and the second order
quantity respectively, and are defined as
\begin{align}
& \frac{\Delta}{a^2}\varphi = \frac{1}{c^2} \left( - 4 \pi G \rho
\delta + \frac{H}{a}\nabla\cdot\mathbi{u} \right) \, ,
\\
& X_2 = 2\varphi \nabla\cdot\mathbi{u} - (\mathbi{u}\cdot\nabla)\varphi +
\frac{3}{2}\Delta^{-1}\nabla\cdot \left[ \mathbi{u} \Delta\varphi +
\mathbi{u}\cdot\nabla \left(\nabla\varphi \right) \right] \, .
\end{align}
In relativistic perturbation theory, the dimensionless quantity
$\varphi$ is proportional to the spatial curvature perturbation in
the comoving gauge.
All the perturbation variables $\delta$, $\mathbi{u}$ and $\varphi$ can be regarded as equivalently
gauge-invariant combinations to non-linear order. Proper choice of variables and gauge conditions are important to have these equations. Note that the
relativistic continuity and Euler equations coincide with those from
Newtonian fluid approximation up to the second order in
perturbations~\citep{peebles:1980,noh/hwang:2004}. Therefore, the perturbative
solutions are also the same up to the second order, and pure general
relativistic effects appear from third order. We emphasize that the
above equations are valid in the presence of the cosmological
constant in the background world model.

An examination of the third order terms in Equations\
(\ref{delta-dot-eq}) and (\ref{u-dot-eq}) shows that the pure third
order terms are simple convolutions of the linear order $\varphi$
with the second-order combinations of fluid variables $\delta$ and
${\mathbi u}$. Note that to the linear order, $\varphi$ is a well-known
conserved quantity whose amplitude of the growing mode solution is
conserved on the super-horizon scales, independent of the changing
equation of state or even changing underlying gravity theories~\citep{hwang/noh:2005a}.
In a flat background without cosmological constant the
amplitude of $\varphi$ near horizon scale is directly related to the
amplitude of relative temperature fluctuations of the CMB as
$\delta T/T=\varphi/5$.

The linear solutions of Equations\ (\ref{delta-dot-eq}) and (\ref{u-dot-eq})
are easily found to be
\begin{align}
\delta_1({\mathbi k},t) = & D(t)\delta_1({\mathbi k},t_0) \, ,
\\
\theta_1({\mathbi k},t) = & -aHD(t)\delta_1({\mathbi k},t_0) \, ,
\end{align}
where $D(t)$ is the linear growth factor so that $\delta_1({\mathbi k},t_0)$ is
the present linear density fluctuation. With these linear solutions, we can
perturbatively expand the density contrast $\delta({\mathbi k},t) = \delta_1
+ \delta_2 + \delta_3 + \cdots$, where $\delta_n$ is a $n$-th order quantity
in linear density contrast $\delta_1({\mathbi k},t_0)$, and similarly for $\theta({\mathbi k},t)$. With this expansion,
we can find the full non-linear solutions of Equations\ (\ref{delta-dot-eq})
and (\ref{u-dot-eq}) by using momentum dependent symmetric kernels as
\begin{align}
\label{deltasol}
\delta({\mathbi k},t) = & \sum_{n=1}^\infty D^n(t) \int \frac{d^3q_1\cdots d^3q_n}{(2\pi)^{3(n-1)}} \delta^{(3)}\left( {\mathbi k}-\sum_{i=1}^n{\mathbi q}_i \right)
\nonumber\\
& \hspace{4em} \times F_n^{(s)}({\mathbi q}_1,\cdots{\mathbi q}_n) \delta_1({\mathbi q}_1)\cdots\delta_1({\mathbi q}_n) \, ,
\\
\label{thetasol}
\theta({\mathbi k},t) = & -aH\sum_{n=1}^\infty D^n(t) \int \frac{d^3q_1\cdots d^3q_n}{(2\pi)^{3(n-1)}} \delta^{(3)}\left( {\mathbi k}-\sum_{i=1}^n{\mathbi q}_i \right)
\nonumber\\
& \hspace{6em} \times G_n^{(s)}({\mathbi q}_1,\cdots{\mathbi q}_n) \delta_1({\mathbi q}_1)\cdots\delta_1({\mathbi q}_n) \, .
\end{align}
Then, Equations\ (\ref{delta-dot-eq}) and (\ref{u-dot-eq}) become simple
differential equations of $F_n^{(s)}$ and $G_n^{(s)}$. Especially, the general
relativistic terms which explicitly include $k_H \equiv aH$, the comoving
wavenumber corresponding to the comoving horizon, are reduced to the algebraic
equations
\begin{align}
& 2F_{3,\text{Einstein}} - G_{3,\text{Einstein}} = -\frac{5}{2}k_H^2 \bigg\{2\frac{{\mathbi q}_1\cdot{\mathbi q}_3}{q_1^2q_2^2}
\nonumber\\
& + \frac{{\mathbi q}_{12}\cdot{\mathbi q}_3}{q_{12}^2} \bigg[ -\frac{2}{q_2^2} + \frac{{\mathbi q}_1\cdot{\mathbi q}_2}{q_1^2q_2^2} - \frac{3}{2}\frac{{\mathbi q}_{12}\cdot{\mathbi q}_1}{q_{12}^2q_1^2} - \frac{3}{2}\frac{{\mathbi q}_{12}\cdot{\mathbi q}_2}{q_{12}^2} \frac{{\mathbi q}_1\cdot{\mathbi q}_2}{q_1^2q_2^2} \bigg] \bigg\} \, ,
\\
& \frac{3}{2}F_{3,\text{Einstein}} - \frac{5}{2}G_{3,\text{Einstein}} = -\frac{5}{2}k_H^2 \bigg\{ \bigg[ \frac{2}{3} + \frac{{\mathbi q}_1\cdot{\mathbi q}_{23}}{q_1^2} \left( 1 - \frac{k^2}{q_{23}^2} \right) \bigg]
\nonumber\\
& \hspace{0.5em} \times \bigg[ -\frac{2}{q_3^2} + \frac{{\mathbi q}_2\cdot{\mathbi q}_3}{q_2^2q_3^2} - \frac{3}{2}\frac{{\mathbi q}_{23}\cdot{\mathbi q}_2}{q_{23}^2q_2^2} - \frac{3}{2}\frac{{\mathbi q}_{23}\cdot{\mathbi q}_3}{q_{23}^2} \frac{{\mathbi q}_2\cdot{\mathbi q}_3}{q_2^2q_3^2} \bigg]
\nonumber\\
& + \frac{1}{q_3^2} \bigg[ \frac{2}{3}\frac{{\mathbi q}_1\cdot{\mathbi q}_2}{q_2^2} - 4 \left( \frac{{\mathbi q}_1\cdot{\mathbi q}_2}{q_2^2} - \frac{1}{3} \right) \frac{{\mathbi k}\cdot{\mathbi q}_1}{q_1^2} \bigg] \bigg\} \, ,
\end{align}
where we have introduced ${\mathbi q}_{12\cdots n} = \sum_{i=1}^n{\mathbi q}_i$.
The second and third order Newtonian kernels can be found in
e.g. Equations [(2.32), (2.33)] and [(2.34), (2.35)] in \citet{jeong:2010},
respectively.

\section{Matter and velocity power spectra}
\label{sec:spectrum}

From Equation\ (\ref{deltasol}), we can find the non-linear power spectrum, which is defined as
\begin{equation}\label{Pdefinition}
\left\langle \delta({\mathbi k}_1,t) \delta({\mathbi k}_2,t) \right\rangle
\equiv (2\pi)^3 \delta^{(3)}({\mathbi k}_1+{\mathbi k}_2)P(k_1,t) \, .
\end{equation}
If we assume perfect Gaussianity of $\delta_1$, which is a very good
approximation consistent with current observations, any higher order
correlation function beyond the linear power spectrum $P_{11}(k)$
disappears and $P_{11}(k)$ is all that we need to specify the
statistics of density fluctuation $\delta$:
as we will see shortly, all the non-linear corrections to the power spectrum
can be written in terms of $P_{11}$.
Then, from
Equation\ (\ref{Pdefinition}) we can write, beyond the linear density power
spectrum $P_{11}$,
\begin{equation}
P = P_{11} + P_{22} + P_{13} + \cdots \, ,
\end{equation}
with $\left\langle\delta_i(\mathbi{k}_1)\delta_j(\mathbi{k}_2)\right\rangle
\equiv (2\pi)^3 \delta^{(3)}(\mathbi{k}_1+\mathbi{k}_2) s_{ij} P_{ij}(k_1)$. Here,
$s_{ij}$ is a symmetric factor which is 1 for $i=j$ and $1/2$
otherwise. The leading non-linear
correction $P_{12}$ includes bispectrum and thus
disappears according to our assumption of Gaussianity of $\delta_1$.
$P_{22} + P_{13}$ denotes the next-to-leading order non-linear correction to
the power spectrum. As mentioned above, $P_{13}$ includes general
relativistic terms.

The density power spectrum
up to next-to-leading order non-linear corrections is
\begin{align}\label{Ptotal}
\nonumber
&P(k,t) =  P_{11}(k,t)
+ \frac{1}{98}\frac{k^3}{(2\pi)^2} \int_0^\infty dr P_{11}(kr,t)
\\
&\times
\int_{-1}^1 dx P_{11}\left( k\sqrt{1+r^2-2rx},t \right) \frac{\left( 3r+7x-10rx^2 \right)^2}{\left( 1+r^2-2rx \right)^2}
\nonumber\\
+ &\frac{1}{252}\frac{k^3}{(2\pi)^2} P_{11}(k,t)
\int_0^\infty dr P_{11}(kr,t)
\nonumber
\\
\nonumber
&\times
\left[ -42r^4 + 100r^2 - 158 + \frac{12}{r^2} + \frac{3}{r^3} \left( r^2-1 \right)^3 \left( 7r^2+2 \right) \log \left| \frac{1+r}{1-r} \right| \, \right]
\nonumber\\
\nonumber
+& \frac{5}{56} \left( \frac{k_H}{k} \right)^2
\frac{k^3}{(2\pi)^2} P_{11}(k,t)
\int_0^\infty dr P_{11}(kr,t)
\\
\nonumber
&\times
\left[ 86r^2 - 130 - \frac{72}{r^2} + \frac{1}{r^3} \left( 36 + 53r^2 - 46r^4 - 43r^6 \right) \log \left| \frac{1+r}{1-r} \right| \, \right]
\nonumber\\
\equiv & P_{11} + P_{22} + P_{13,\rm Newton} + P_{13,\rm Einstein}
\, .
\end{align}
where $r$ and $x$ are the magnitude of dummy integration momentum $\mathbi q$ and the cosine between $\mathbi q$ and $\mathbi k$, respectively, introduced as $q \equiv rk$ $(0 \leq r \leq \infty)$ and ${\mathbi k}\cdot{\mathbi q} \equiv k^2rx$ $(-1 \leq x \leq 1)$.
We have divided $P_{13}$ into the Newtonian part $P_{13,\rm Newton}$
and the general relativistic contribution $P_{13,\rm Einstein}$.
Compared with
$P_{13,\rm Newton}$ the general relativistic contribution
$P_{13,\rm Einstein}$ is multiplied by a factor $(k_H/k)^2$, where $k_H/k$ is the ratio between a scale of interest and the horizon scale, and is thus
highly suppressed far inside the horizon.

\begin{figure}[h]
\begin{center}
\rotatebox{90}{
\includegraphics[width=6.5cm]{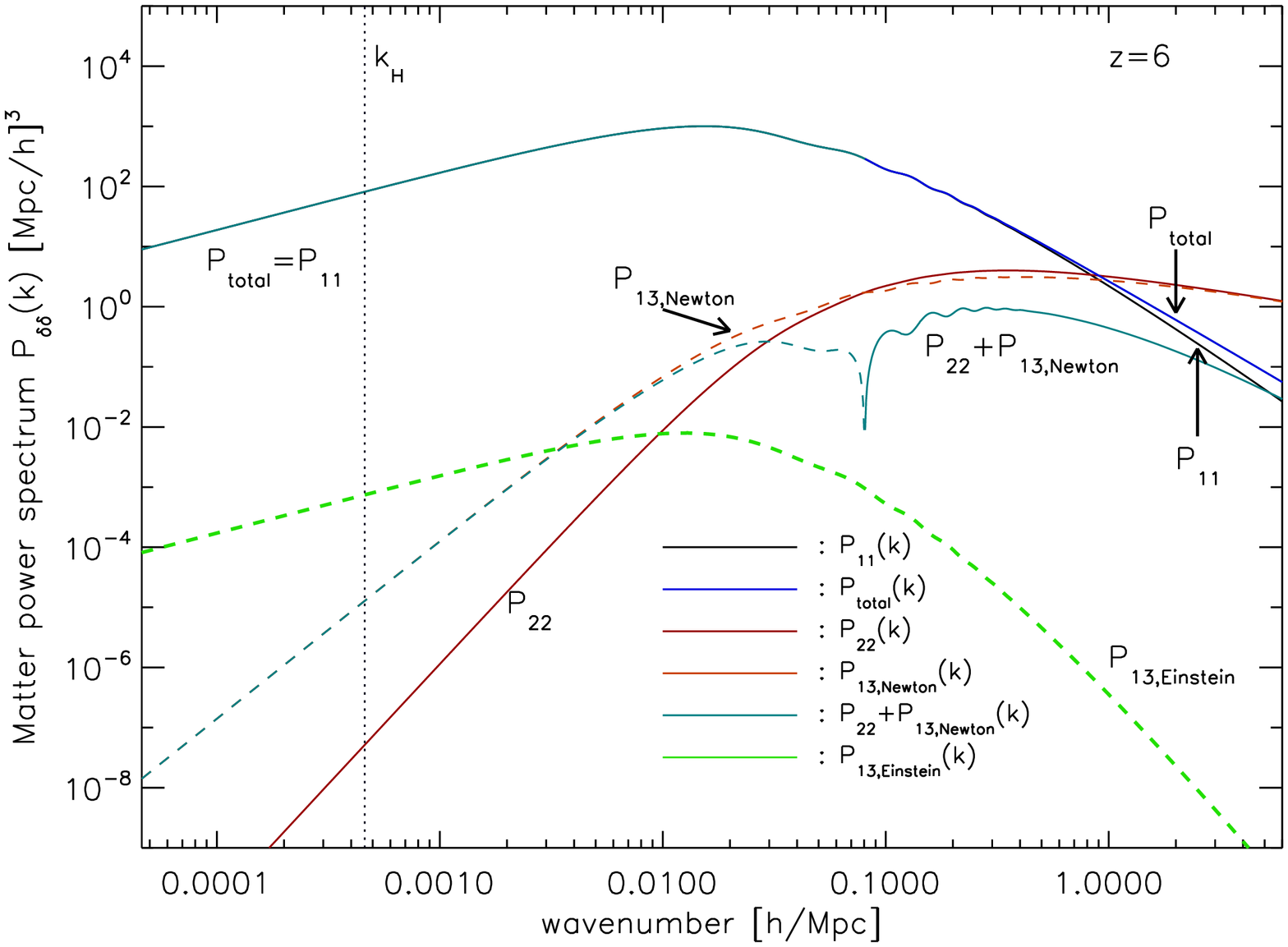}
}
\end{center}
 \caption{
         Non-linear matter power spectrum (solid blue line)
         and the contribution from each
         component of Equation\ (\ref{Ptotal}) at $z=6$.
         The black, red, and orange lines show the contributions
            from the Newtonian perturbation theory:
         $P_{11}$, $P_{22}$, and $P_{13,\rm Newton}$,
         respectively.
         The green line shows the general relativistic
         effect, $P_{13,\rm Einstein}(k)$.
         Note that we take the absolute value
         for negative terms, and show with dashed lines:
         $P_{22}$ and $P_{13}$ are positive and negative,respectively, in all scales.
         Vertical dotted line
         shows the wavenumber correspond to the comoving horizon $k_H$ at
            $z=6$. Over all scales, the general relativistic term
            $P_{13,\rm Einstein}$ (green)
            is negligibly small compare to the linear
            power spectrum $P_{11}$ (black).
}
 \label{fig:Ptotal}
\end{figure}

In Figure~\ref{fig:Ptotal} we present the total power spectrum of
Equation\ (\ref{Ptotal}) along with its components $P_{11}$, $P_{22}$,
$P_{13,\rm Newton}$ and $P_{13,\rm Einstein}$ when our Universe is
dominated by matter, at $z=6$. The linear power spectrum is
calculated by {\sf CAMB}~\citep{lewis/challinor/lasenby:2000} code with the maximum
likelihood cosmological parameters given in the Table 1 of
\citet{komatsu/etal:2008} (``WMAP+BAO+SN'').
Figure~\ref{fig:Ptotal} shows that the general relativistic
contribution $P_{13,\rm Einstein}$ is smaller than the linear power
spectrum $P_{11}$ on {\em all} cosmological scales.

Let us examine $P_{13,\rm Einstein}$ more closely. For notational
simplicity, we shall abbreviate the integration in $P_{13,\rm
Einstein}$ as $\int dr P_{11}(kr,t)f(r)$. Then, scale dependence of
$P_{13,\rm Einstein}$ can be understood as follows. First, setting
$kr=q$, we find that $P_{13,\rm Einstein} \sim P_{11}(k)f(q/k)$.
On small scales ($k \gg 0.01h/\mathrm{Mpc}$), $q/k$ is also small,
and by using Taylor expansion of $f(r)=-(656/15)r^2 +
\mathcal{O}(r^4)$ we find $P_{13,\rm Einstein} \sim k^{-2} P_{11}$.
On the other hands, in large scale limit ($k \ll
0.01h/\mathrm{Mpc}$) where $q/k$ takes larger value,
$f(r)=-752/3+\mathcal{O}(r^{-2})$ and $P_{13,\rm Einstein}$ has a
scale dependence $P_{13,\rm Einstein} \sim P_{11}$. Numerical
calculation reveals that $P_{13,\rm Einstein}$ is smaller than
$P_{11}$ by a factor $10^{-5}$ on large scales.
Our result shows that the leading order non-linear power spectrum is
finite in both infrared and ultraviolet regions\footnote{The previous result
reporting infrared divergence in $P_{13,\rm Einstein}$~\citep{noh/jeong/hwang:2009}
turns out to be due to an incorrect calculation of the
power spectrum: the third order general relativistic kernels
$F_{3,\mathrm{Einstein}}$ has not been fully symmetrized, thus causing
logarithmic infrared divergence in $P_{13,\mathrm{Einstein}}$.}.

We can proceed in the same way to compute the power spectrum of the
peculiar velocity. As Equation \ (\ref{Pdefinition}), we can define
\begin{equation}
\left\langle \theta({\mathbi k}_1,t)\theta({\mathbi k}_2,t) \right\rangle \equiv (2\pi)^3\delta^{(3)}({\mathbi k}_1+{\mathbi k}_2)P_{\theta\theta}(k_1,t) \, ,
\end{equation}
and we can find
\begin{align}
& k_H^{-2} P_{\theta\theta}(k,t) = P_{11}(k,t) + \frac{1}{98}\frac{k^3}{(2\pi)^2} \int_0^\infty dr P_{11}(kr,t)
\nonumber\\
& \times \int_{-1}^1 dx P_{11}\left( k\sqrt{1+r^2-2rx},t \right) \frac{(r-7x+6rx^2)^2}{(1+r^2-2rx)^2}
\nonumber\\
+ & \frac{1}{84}\frac{k^3}{(2\pi)^2}P_{11}(k,t) \int_0^\infty dr P_{11}(kr,t)
\nonumber\\
& \times \left[ -6r^4 + 4r^2 - 82 + \frac{12}{r^2} + \frac{3}{r^3} \left( r^2-1 \right)^3 \left( r^2+2 \right) \log \left| \frac{1+r}{1-r} \right| \, \right]
\nonumber\\
+ & \frac{5}{56} \left( \frac{k_H}{k} \right)^2 \frac{k^3}{(2\pi)^2} \int_0^\infty dr P_{11}(kr,t)
\nonumber\\
& \times \left[ 46r^2 - 50 - \frac{144}{r^2} + \frac{1}{r^3} \left( -23r^6 - 50r^4 + r^2 + 72 \right) \log \left| \frac{1+r}{1-r} \right| \, \right] \, .
\label{eq:ptt}
\end{align}

\begin{figure}[h]
\begin{center}
\rotatebox{90}{
\includegraphics[width=6.5cm]{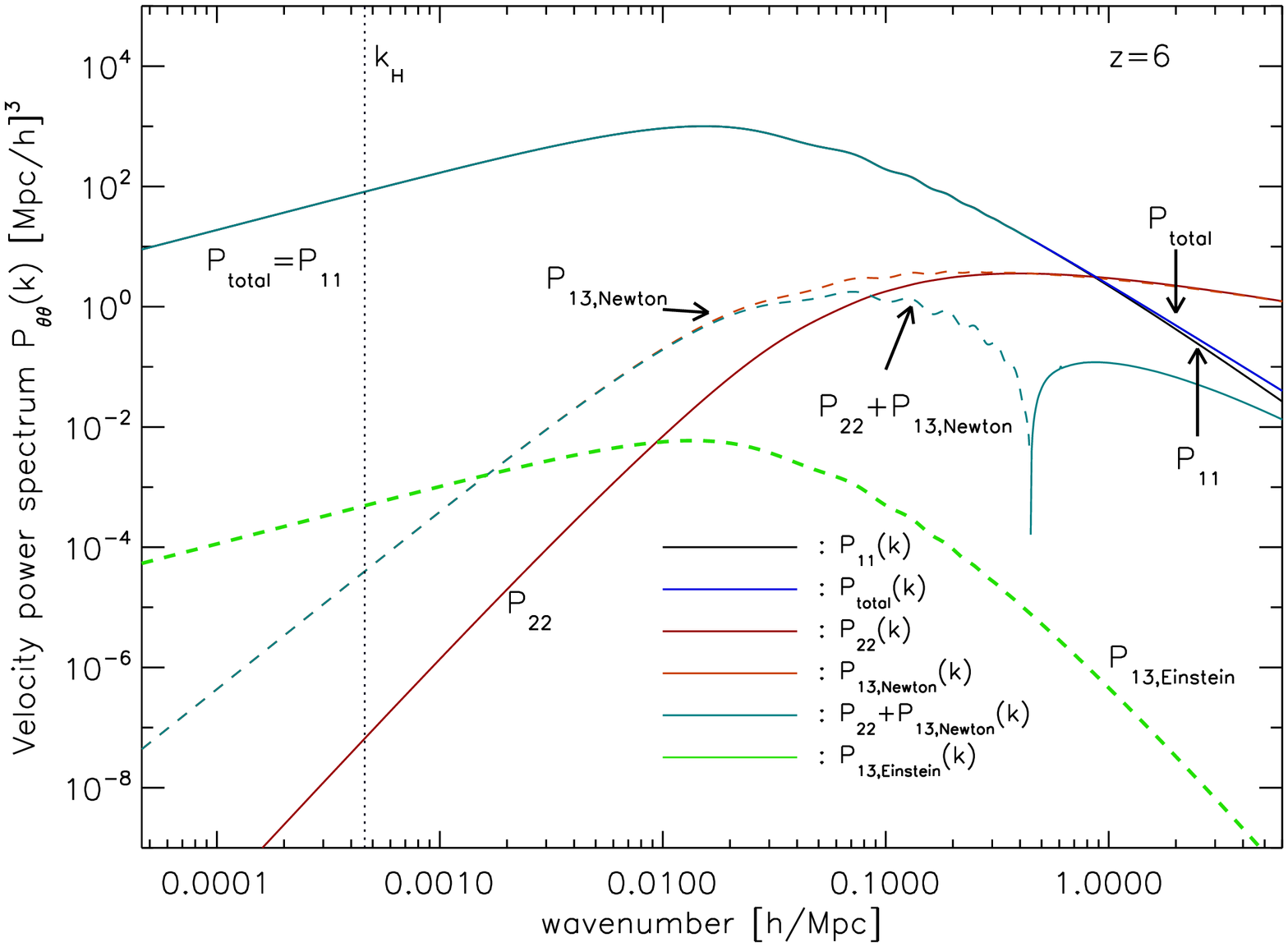}
}
\end{center}
 \caption{
        Same as Figure \ref{fig:Ptotal}, but for the velocity power spectrum
        $P_{\theta\theta}(k)$.
}
 \label{fig:Ptheta}
\end{figure}

Figure \ref{fig:Ptheta} shows the non-linear velocity power spectrum of
Equation\ (\ref{eq:ptt}) for exactly the same cosmology as Figure
\ref{fig:Ptotal}. As in the case of the total matter power spectrum,
the non-linear general relativistic correction is negligibly
small for all scales. It is because the third order kernel for velocity
$G_{3,\mathrm{Einstein}}$ behaves in the same way as that for the matter
density $F_{4,\mathrm{Einstein}}$ in both large ($r\to 0$) and
small ($r\to\infty$) scale limit:
$\lim_{r\to0}g(r)=-(368/15)r^2+\mathcal{O}(r^{4})$
and
$\lim_{r\to\infty}g(r)=-496/3+\mathcal{O}(r^{-2})$ when
denoting the last integration in Equation (\ref{eq:ptt}) as
$\int dr P_{11}(kr)g(r)$.

\section{conclusion}
\label{sec:conclusions}

To conclude, in this note we have examined the general
relativistic non-linear contributions to the density and velocity power spectra.
We have found that, with
pleasant surprise, the pure general relativistic effects are
completely negligible on {\em all} cosmologically relevant scales, even outside the horizon.
It is interesting to see that the linear power spectrum is totally
dominating even outside the horizon. Our conclusion has the
following important implication. As the general relativistic effect
is very small, Newtonian theory can be safely applied to the
non-linear evolution of cosmic structure
on all cosmologically relevant scales.
In the literature it has
been common to use Newtonian gravity to study the non-linear
clustering properties of large scale structure without justifying
that approach. The result we present in this note provides a
confirmation of using Newtonian gravity to handle non-linear
clustering in cosmology.

\acknowledgments

J.G. wishes to thank Misao Sasaki and Takahiro Tanaka for useful
conversations, and is grateful to the Yukawa Institute for
Theoretical Physics, Kyoto University for hospitality during the
workshop YITP-W-10-10 where part of this work was carried out.
This work was supported in part by a
Robinson prize postdoctoral fellowship at
California Institute of Technology (D.J.),
a VIDI and a VICI Innovative Research Incentive Grant from the
Netherlands Organisation for Scientific Research (NWO) (J.G.),
Mid-career Research Program through National Research
Foundation funded by the MEST (No. 2010-0000302) (H.N.)
and
the Korea Research Foundation Grant funded by the Korean
Government (KRF-2008-341-C00022) (J.H.).

\end{document}